\documentclass[%
 reprint,
 amsmath,amssymb,
 aps,
]{revtex4-1}

\usepackage{graphicx}
\usepackage{dcolumn}
\usepackage{bm}

\begin{document}

\preprint{APS/123-QED}

\title{\lq\lq Tiny galaxies\rq\rq~ from \lq\lq little bang\rq\rq}

\author{Feng Li}
\email{fengli@lzu.edu.cn}
\affiliation{School of Physical Science and Technology, Lanzhou University, Lanzhou, Gansu, 073000, China }

\date{\today}

\begin{abstract}
We show, by solving the linearized Vlasov equations, that in the presence of an attractive density-dependent vector, or tensor, mean field potential, under a portion of the spinodal unstable phase region where the attractions among the quarks are more dominant, the transverse modes, i.e., the fluctuations of the current densities perpendicular to the wave vector, are activated due to relativity. In particular, the current densities in both the transverse directions oscillate with the same frequency, indicating the existence of the spontaneous rotations of the dense clusters. These rotating clusters would certainly create a local vortical flow field which, according to chiral (or spin) vortical effect, leads to the macroscopic separation of the axial charges or spins due to the spin-orbital interactions. Hence, for the first time, the connection between the axial charge or spin in-homogeneity and the first-order phase transition is proposed.

\end{abstract}

\pacs{Valid PACS appear here}
\maketitle


Quark gluon plasma (QGP)~\cite{Shuryak:1980tp}, i.e., a novel matter where the quarks and the gluons are liberated from the hadrons, are believed to exist right after the big bang, and re-created in the \lq\lq small bangs\rq\rq, namely, the relativistic heavy ion collisions. The transition from QGP to the normal hadron gas is confirmed, by the lattice calculation~\cite{Bernard:2004je,Aok06,Baz12}, as a cross-over at small baryon chemical potentials, and is expected, by many effective models~\cite{Nam611,Nam612,Asa89,Car10,Bra12,Stephanov:2004wx,Stephanov:2007fk,Fukushima:2008wg,Fukushima:2013rx,Baym:2017whm,Sun:2020bbn}, to be a first-order phase transition at large chemical potentials, while the latter process is driven by the spinodal instability~\cite{Randrup:2010ax,Steinheimer:2012gc,Steinheimer:2013xxa,Li:2015pbv,Li:2016uvu}, and might affect the direct flows~\cite{Rischke:1995pe,Stoecker:2004qu} and the light nuclei yields~\cite{Sun:2017xrx,Sun:2018jhg,Sun:2020pjz}, which are being studied in the beam energy scan program at the Relativistic Heavy Ion Collision (RHIC)~\cite{Adamczyk:2014ipa,Adamczyk:2017nxg,Adam:2019wnb,Zhang:2019wun,Zhang:2020ewj}. The spinodal instability is conventionally characterized by the imaginary speed of sound~\cite{Steinheimer:2012gc}, which corresponds to the non-propagating and self-amplified sound waves, or density fluctuations. Hence, the spinodal unstable modes are usually considered longitudinal, i.e., the current density is in the direction of the wave vector. It will be shown in this work, however, that once the attractions among the quarks are strong enough, the transverse modes will also be activated, due to relativity, under a certain portion of the spinodal unstable phase region. These transverse modes can be interpreted as the spontaneous rotation of the dense clusters. The \lq\lq tiny galaxies\rq\rq, or the rotating clusters, would certainly create a local vortical flow field, which, according to chiral (or spin) vortical effect\cite{Vilenkin:1979ui,Son:2009tf,Kharzeev:2010gr,Jiang:2015cva,Kharzeev:2015znc}, leads to the macroscopic separation of the axial charges or spins due to the spin-orbital interactions. We therefore, for the first time, build the connection between the first order phase transition and the axial charge or spin in-homogeneity of the hadrons observed and measured in RHIC \cite{Abelev:2007zk,STAR:2017ckg,Niida:2018hfw,Adam:2018ivw,Adam:2019srw}.

An attractive mean field potential acting on the quarks is necessary for describing the first-order phase transition of the quark matter. In this work, we adopt a simple density dependent vector mean-field potential which, in the Landau local rest frame, is in form of $U(\rho^\ast)$ depending only on $\rho^\ast$, with $\rho^\ast\equiv\sqrt{j_\mu j^\mu}$ being the net quark density in the Landau local rest frame. Quarks move, under such a mean-field potential, according to the covariant equations of motion~\cite{EOM}, i.e.,
\begin{eqnarray}
    \dot {\mathbf x} &=& \mathbf p/E_p, \nonumber\\
    \dot {\mathbf p} &=& \mp U^\prime(\rho^\ast)\left(\nabla \rho+\partial_t \mathbf j - \dot {\mathbf x} \times (\nabla\times\mathbf j)\right),
    \label{eq:eom}
\end{eqnarray}
where $\mathbf p$ denotes the kinetic momentum, and $\mp$ is taken \lq-\rq~ for the quarks, and \lq+\rq~ for the anti-quarks. Given the negligible contribution from the particle scatterings, the phase-space distribution function $f_\pm(\mathbf x, \mathbf p, t)$, with the subscription $\pm$ denoting the particles and anti-particles respectively, hence evolves according to the Vlasov equations, i.e.,
\begin{eqnarray}
    0 &=& \partial_t f_\pm + \mathbf v_{\mathbf p}\cdot \nabla f_\pm \nonumber\\
    && \mp U^\prime(\rho^\ast)\left(\nabla \rho+\partial_t \mathbf j - \mathbf v_{\mathbf p} \times (\nabla\times\mathbf j)\right)\cdot \nabla_{\mathbf p} f_\pm,
    \label{eq:Vlasov}
\end{eqnarray}
where $\mathbf v_{\mathbf p}\equiv \mathbf p/E_p$. Expand $f(\mathbf x,\mathbf p, t)$ as $f_0(E_p)+\delta f(\mathbf x, \mathbf p, t)$ with $f_0(E_p)$ being the equilibrium quark distribution in a stationary and uniform system, i.e., Fermi-Dirac distribution, and keep only the leading terms of $\delta f$, one obtain that
\begin{equation}
    (\omega-\mathbf v_{\mathbf p}\cdot \mathbf k)\delta \widetilde{f}_\pm\pm U^\prime(\rho^\ast)f^\prime_{0\pm}(E_p)\mathbf v_{\mathbf p}\cdot\left(\mathbf k \delta \widetilde \rho - \omega\delta\widetilde{\mathbf j}\right)=0,
\end{equation}
where $\delta\widetilde{f}_\pm(\mathbf k, \mathbf p, \omega)$
\begin{equation}
    \delta\widetilde{f}_\pm(\mathbf k, \mathbf p, \omega)\equiv\int d^4 x \delta f_\pm(\mathbf x, \mathbf p, t) e^{i\omega t-i\mathbf k\cdot \mathbf x}
\end{equation}
is the Fourier transformed distribution fluctuation, and $\delta\widetilde\rho$ and $\delta\widetilde{\mathbf j}$ are related to $\delta\widetilde f_\pm$ by
\begin{eqnarray}
    \delta \widetilde\rho &=& N_c\int \frac{d^3\mathbf p}{(2\pi)^3} (\delta \widetilde f_+ - \delta \widetilde f_-), \\
    \delta \widetilde{\mathbf j} &=& N_c\int \frac{d^3\mathbf p}{(2\pi)^3} \mathbf v_{\mathbf p} (\delta\widetilde f_+ - \delta\widetilde f_-).
\end{eqnarray}
After transforming $\delta \widetilde f_\pm$ to $\delta\widetilde\rho$ and $\delta\widetilde{\mathbf j}$ by integrating over the momentum space, we obtain a group of homogeneous linear equations about $\delta\widetilde\rho$ and $\delta\widetilde{\mathbf j}$, i.e.,
\begin{eqnarray}
    \label{eq:SoundMode}
    \left[I-U^\prime(\rho^\ast)\chi_L(\omega,\mathbf k)\right]
    \left(
    \begin{array}{c}
         \delta\widetilde\rho  \\
         \delta\widetilde j_z 
    \end{array}
    \right) &=& 0, \\
    \label{eq:TransMode}
    \left[ 1-U^\prime(\rho^\ast)\chi_T(\omega,\mathbf k) \right] \delta j_{x,y} &=& 0,
\end{eqnarray}
where
\begin{equation}
    \chi_L \equiv \left(
    \begin{array}{cc}
    - |k| \chi^{(1)}_L(\omega,\mathbf k)    &   \omega \chi^{(1)}_L(\omega,\mathbf k)\\
    - |k| \chi^{(2)}_L(\omega,\mathbf k)     &   \omega \chi^{(2)}_L(\omega,\mathbf k)
    \end{array}
    \right),
\end{equation}
with the response functions $\chi^{(1,2)}_L$ and $\chi_T$ being
\begin{eqnarray}
    \chi^{(1)}_L &=& N_c\int\frac{d^3\mathbf p}{(2\pi)^3}\frac{\mathbf v_{\mathbf p}\cdot\hat{\mathbf k}}{\omega-\mathbf v_{\mathbf p}\cdot\mathbf k} \left(f_{0+}^\prime(E_p)+f_{0-}^\prime(E_p)\right),\\
    \chi^{(2)}_L &=& N_c\int\frac{d^3\mathbf p}{(2\pi)^3}\frac{(\mathbf v_{\mathbf p}\cdot\hat{\mathbf k})^2}{\omega-\mathbf v_{\mathbf p}\cdot\mathbf k} \left(f_{0+}^\prime(E_p)+f_{0-}^\prime(E_p)\right),\\
    \chi_T &=& \frac{\omega N_c}{2}\int\frac{d^3\mathbf p}{(2\pi)^3}\frac{\mathbf v_{\mathbf p}^2-(\mathbf v_{\mathbf p}\cdot\hat{\mathbf k})^2}{\omega-\mathbf v_{\mathbf p}\cdot\mathbf k} \left(f_{0+}^\prime(E_p)+f_{0-}^\prime(E_p)\right),\nonumber\\
\end{eqnarray}
and $z-$axis being chosen in the $\mathbf k$ direction.
Note that Eq.(\ref{eq:TransMode}), describing the evolution of the fluctuations of the transverse current density, comes from the \lq\lq$\partial_t\mathbf{j}$\rq\rq~ term in Eq.(\ref{eq:Vlasov}), and hence a consequence of the relativity.

\begin{figure}
    \includegraphics[width=0.48\textwidth]{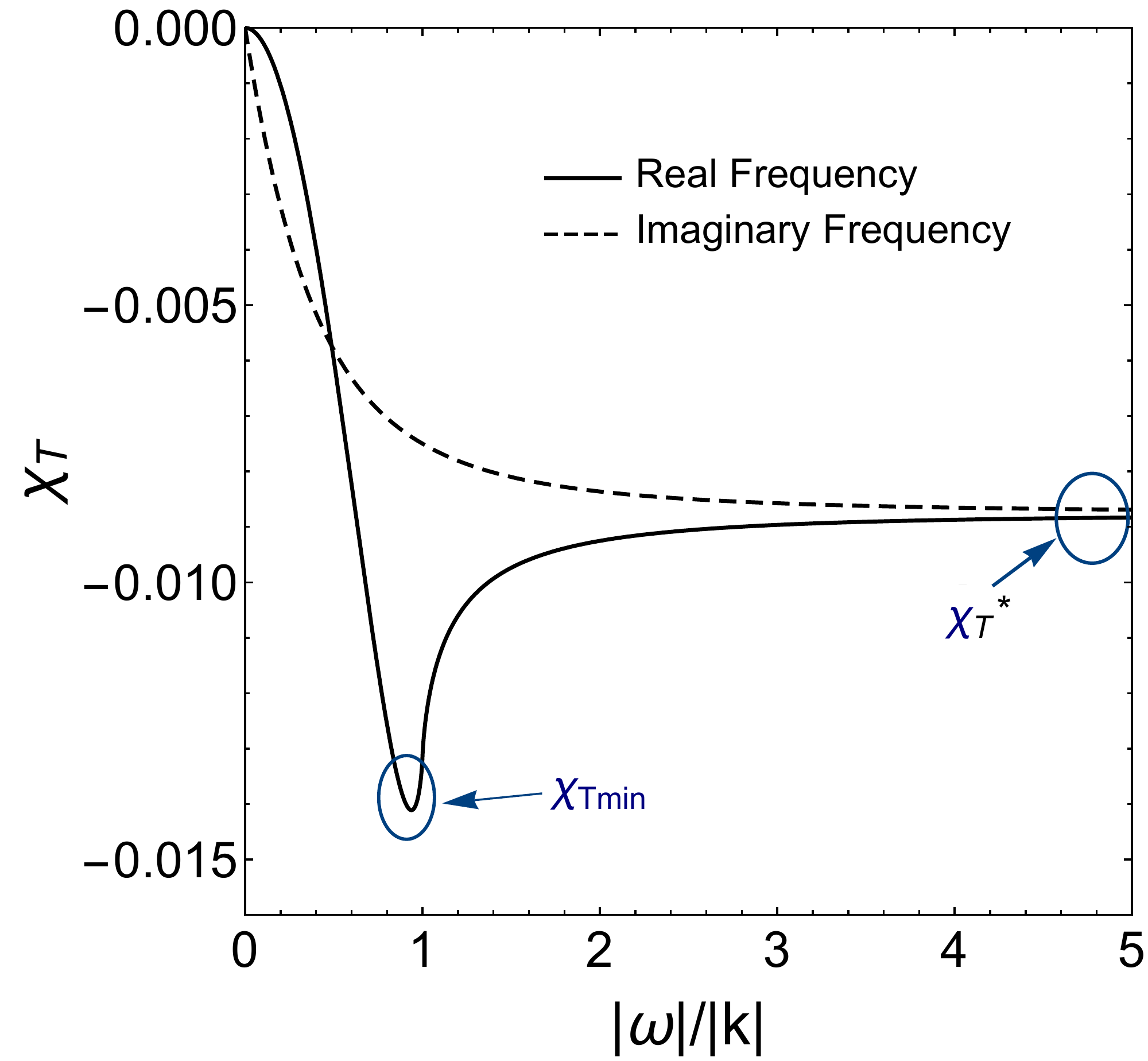}
    \caption{The transverse response functions $\chi_T(\omega,k)$ with the frequency $\omega$ being purely real (represented by solid line) and imaginary (represented by dashed line) obtained at $T=50$ MeV and $n=0.45$ fm$^{-3}$.}
    \label{fig:Chi_T}
\end{figure}
A trivial solution to Eq.(\ref{eq:SoundMode},\ref{eq:TransMode}) is that $0=\delta \rho = \delta\mathbf j$ corresponding to the cases with vanishing fluctuations. The non-vanishing fluctuations are plausible if, and only if $\mathrm{det}(I-U^\prime(\rho^\ast)\chi_{L/T})=0$, which gives the eigen-frequencies of both the longitudinal, or the sound, mode and the transverse modes, respectively. The eigen-frequencies of the sound mode have been extensively studied under various models~\cite{Randrup:2010ax,Steinheimer:2012gc,Li:2015pbv}. It is easy to prove that, in the presence of the current mean-field potential, the eigen-frequency of a longitudinal mode would be imaginary, indicating the existence of a spinodal unstable mode, if and only if
\begin{equation}
\label{eq:SpinodalCriterion}
    U^\prime(\rho^\ast) < \chi^{\ast-1}_L
\end{equation}
where
\begin{eqnarray}
    \chi^{\ast}_L &=& N_c\int\frac{d^3\mathbf p}{(2\pi)^3} \left(f_{0+}^\prime(E_p)+f_{0-}^\prime(E_p)\right)\nonumber\\
    &=& -(\partial \rho^\ast /\partial \mu^\ast)_T
\end{eqnarray}
being negatively valued with $\mu^\ast\equiv\mu-U$ being the effective chemical potential. It is straightforward that Eq.(\ref{eq:SpinodalCriterion}) is equivalent to the isothermal spinodal criterion, i.e., $(\partial p/\partial \rho^\ast)_T<0$. Meanwhile, it can be shown, by integrating over the solid angles, that $\chi_L^\ast < \chi_T(\omega, k)$ regardless how large the $\omega$ or $k$ is at any temperatures or chemical potentials.

The eigen-frequencies of the transverse modes, given by
\begin{equation}
    \label{eq:EigenTransverse}
    U^\prime(\rho^\ast)\chi_T(\omega, k) = 1,
\end{equation}
must be either purely real or purely imaginary for keeping the left hand side of Eq.(\ref{eq:EigenTransverse}) real. In the former case, both $\delta j_x$ and $\delta j_y$ oscillate with the same frequency, which leads to a transverse flow circling either clockwise or anti-clockwise, depending on the relative phase between $\delta j_x$ and $\delta j_y$, while in the latter case, given the evenness of $\chi_T$ in $\omega$, both $\delta j_x$ and $\delta j_y$ grow exponentially with time with the same growth rate. It is apparent that $\chi_T(\omega, k)$ depends only on $\omega/k$, indicating that, in the classical limit, the angular frequency of the circling transverse flow is proportional to the wave number, or inversely proportional to the wavelength, i.e., the longitudinal size of the cluster. Quantum corrections, as mentioned in Ref.~\cite{Li:2015pbv}, might modify the dispersion relation at the large $k$ limit, which is out of the scope of this study.

\begin{figure}
    \includegraphics[width=0.48\textwidth]{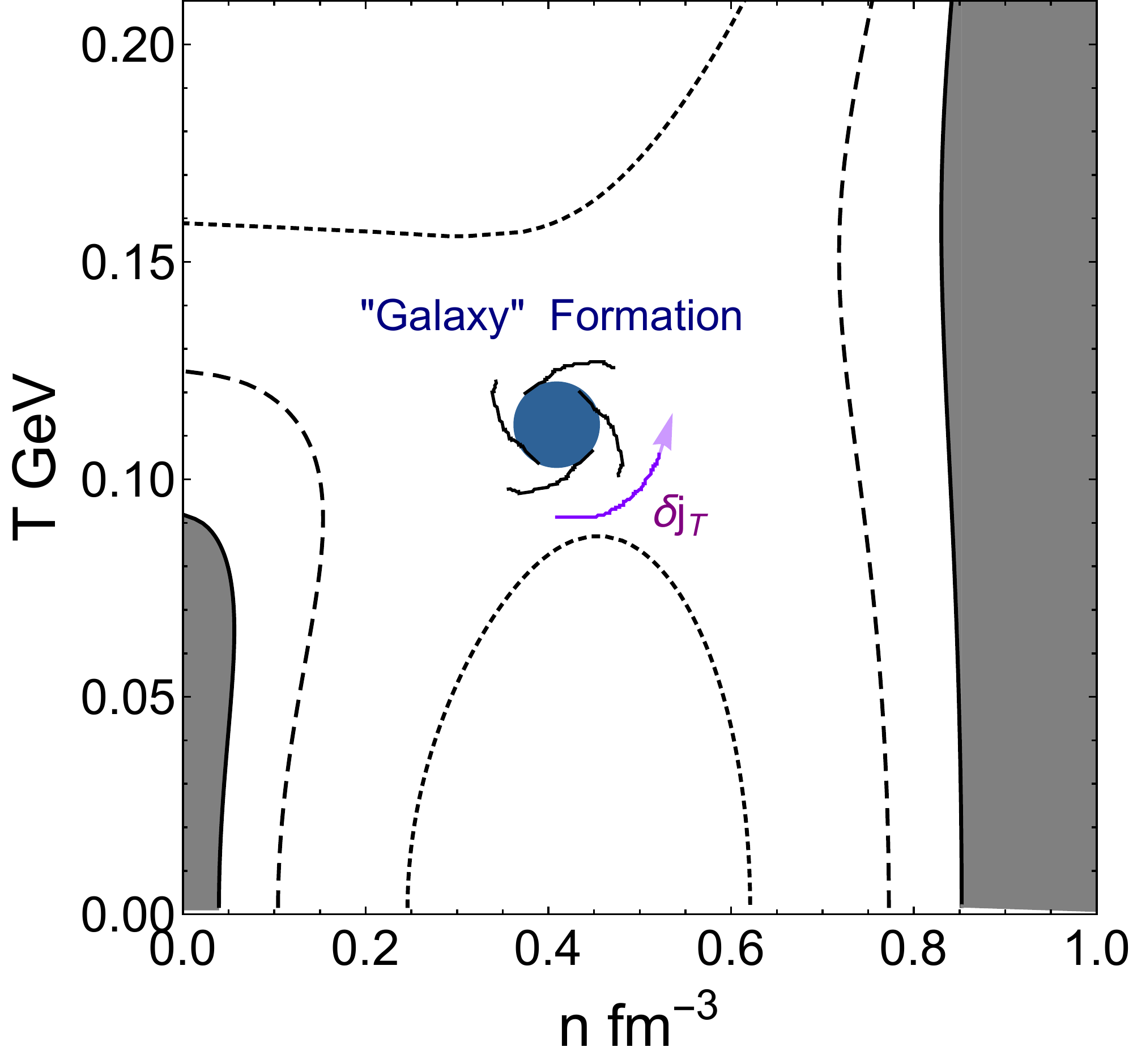}
    \caption{The phase regions obtained in the presence of the mean-field potential $U(\rho^\ast)=A(\rho^\ast/\rho_0)+B(\rho^\ast/\rho_0)^\sigma$} with $A$, $B$, $\sigma$, $\rho_0$ and the quark mass $m$ being $-700$ MeV, $100$ MeV, $2.6$, $0.5$ fm$^{-3}$, and $6$ MeV, respectively. The spinodal unstable phase region is colored in white, and the boundaries of the phase regions, where the transverse modes are activated and self-amplified, are marked by both the long and short dashed lines, respectively. The dense clusters spontaneously rotate under the phase region between the long and short dashed boundaries.
    \label{fig:PhaseDiagram}
\end{figure}
However, the solution to Eq. (\ref{eq:EigenTransverse}) does not always exist, since, as $\omega/k$ varies on either the whole real or the whole imaginary axes, the value of $\chi_T(\omega/k)$ varies only in a finite range. As an example, we plot $\chi_T(\omega/k)$ with the quark mass being 6 MeV at $T=50$ MeV and $\rho^\ast=0.45$ fm$^{-3}$ for both the cases with $\omega/k$ being both purely real and imaginary in Fig. \ref{fig:Chi_T} by the solid and the dashed lines, respectively. It is shown that, in both the cases, $\chi_T(0)=0$ and $\lim_{\omega/k\to\infty}\chi_T(\omega/k)=\chi^\ast_T$, where
\begin{equation}
    \chi_T^\ast = \frac{N_c}{3} \int\frac{d^3\mathbf p}{(2\pi)^3} v_{\mathbf p}^2 \left(f_{0+}^\prime(E_p)+f_{0-}^\prime(E_p)\right)
\end{equation}
being negatively valued as well. For the cases with the frequencies being purely imaginary, the transverse response function should satisfy $\chi_T \in (\chi_T^\ast,0)$, while for the cases with the frequencies being purely real, the minimum value of the response function $\chi_{T\rm{min}}$, being smaller than $\chi_T^\ast$, is achieved at about $\omega/k\approx 0.94$. All these features are robust against the variation of the temperature and the net quark density. We therefore conclude that Eq. (\ref{eq:EigenTransverse}) must have one purely real solution, labelled as $\omega_1$ and one purely imaginary solution, labelled as $i\Gamma$, as long as
\begin{equation}
    \label{eq:TransverseCriterion2}
    U^\prime(\rho^\ast) < \chi^{\ast-1}_T
\end{equation}
is fulfilled, and must have two purely real solutions, labelled as $\omega_1$ and $\omega_2$ for the lower and higher frequencies, respectively, as long as 
\begin{equation}
    \label{eq:RotationCriterion}
    \chi^{\ast-1}_T < U^\prime(\rho^\ast) < \chi^{-1}_{T\rm{min}}
\end{equation}
is fulfilled. Given the existence of imaginary root to Eq. (\ref{eq:EigenTransverse}), the exponential self-amplification would dominate the time evolution of both $\delta j_x$ and $\delta j_y$ if Eq. (\ref{eq:TransverseCriterion2}) is fulfilled. So, although 
\begin{equation}
    \label{eq:TransverseCriterion}
    U^\prime(\rho^\ast) < \chi^{-1}_{T\rm{min}}
\end{equation}
is the sufficient and necessary criterion for the activation of the transverse modes, Eq. (\ref{eq:RotationCriterion}) is the criterion for the occurrence of the dense cluster rotation. Since, as mentioned in the previous section, $\chi_{T\rm{min}}$ is larger than $\chi^\ast_L$, Eq.  (\ref{eq:SpinodalCriterion}) must hold true if Eq. (\ref{eq:RotationCriterion}) is fulfilled, which means that the phase region, under which the dense clusters spontaneously rotate, is enclosed by the spinodal unstable phase region. Although the transverse modes are activated under the phase region where $U^\prime(\rho^\ast)$ is more negative, indicating stronger attractions among the quarks, it does not mean that the clusters rotate faster if the attractions are stronger. In fact, the lower angular frequency $\omega_1$ increases with $U^\prime(\rho^\ast)$ as $U^\prime(\rho^\ast)\in (-\infty,\chi^{-1}_{T\rm{min}})$, indicating a faster rotation occurring in the case where the attractions are weaker, while the higher angular frequency $\omega_2$ decreases with $U^\prime(\rho^\ast)$ as $U^\prime(\rho^\ast)\in (\chi^{\ast-1}_T,\chi^{-1}_{T\rm{min}})$, indicating a faster rotation occurring in the case where the attractions are stronger. The growth rate of the transverse mode $\Gamma$ is larger when the attractions are weaker as $U^\prime(\rho^\ast)\in (-\infty,\chi^{\ast-1}_T)$ as well.

\begin{figure}
    \includegraphics[width=0.44\textwidth]{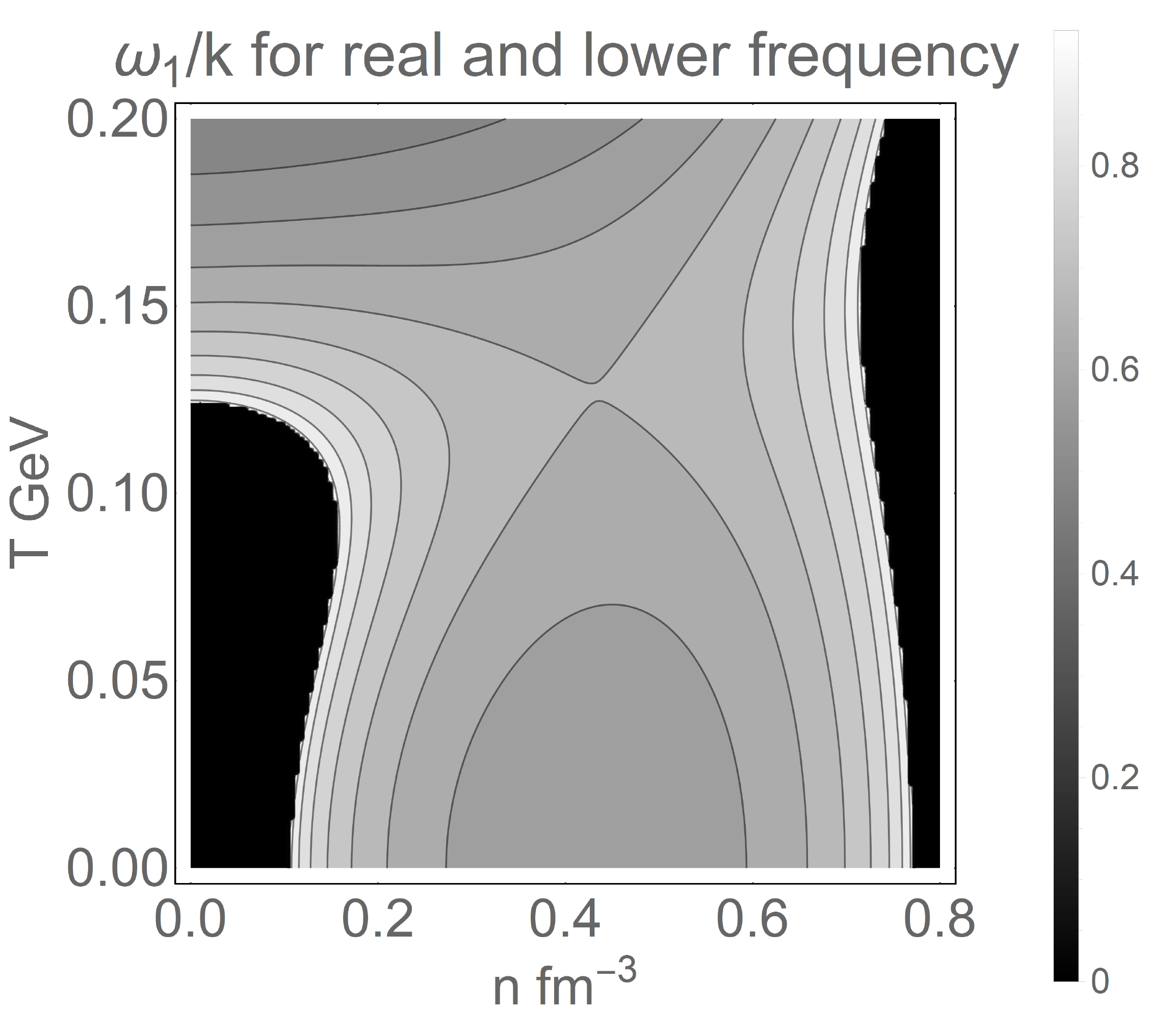}
    \includegraphics[width=0.44\textwidth]{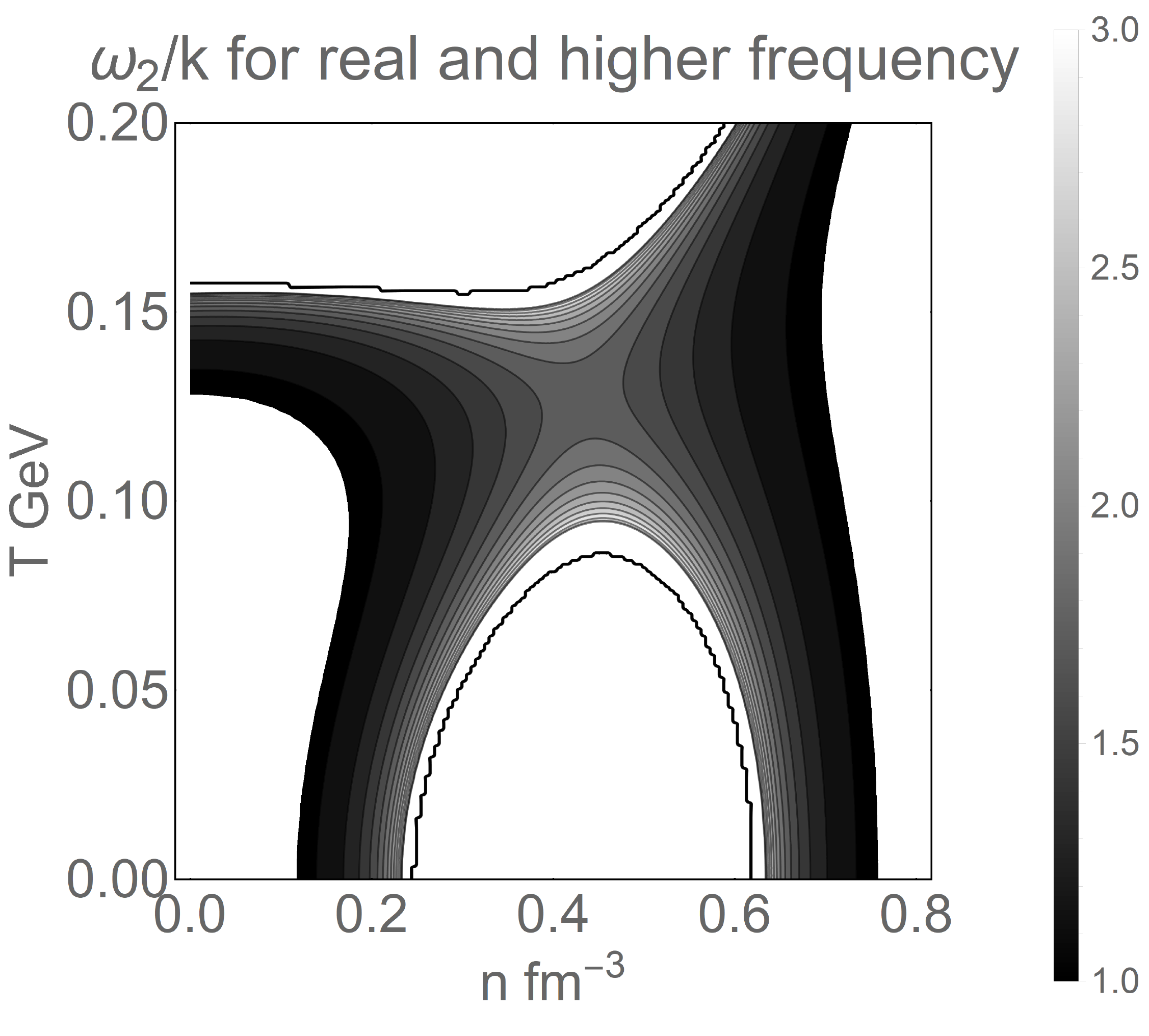}
    \includegraphics[width=0.44\textwidth]{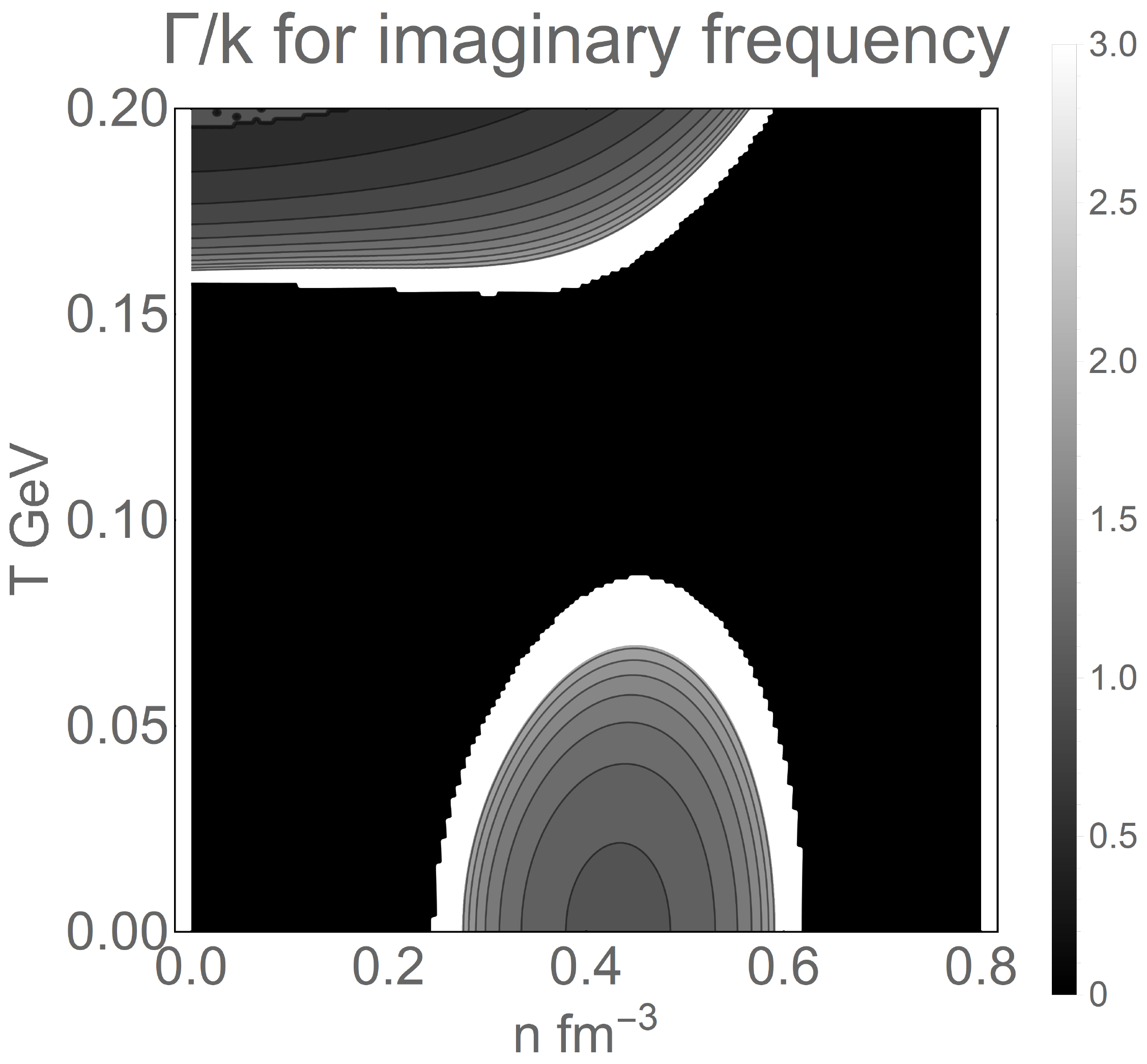}
    \caption{The lower frequency, higher frequency and growth rate, divided by the wave number, obtained in the presence of the mean-field potential $U(\rho^\ast)=A(\rho^\ast/\rho_0)+B(\rho^\ast/\rho_0)^\sigma$} with $A$, $B$, $\sigma$, $\rho_0$ and the quark mass $m$ being $-700$ MeV, $100$ MeV, $2.6$, $0.5$ fm$^{-3}$, and $6$ MeV, respectively.
    \label{fig:Freqencies}
\end{figure}
As an example, we show, in Fig. \ref{fig:PhaseDiagram}, the phase regions obtained under the mean-field potential
\begin{equation}
    U(\rho^\ast)=A\left(\frac{\rho^\ast}{\rho_0}\right)+B\left(\frac{\rho^\ast}{\rho_0}\right)^\sigma
\end{equation}
with $A$, $B$, $\sigma$, $\rho_0$ and the quark mass $m$ being $-700$ MeV, $100$ MeV, $2.6$, $0.5$ fm$^{-3}$, and $6$ MeV, respectively, by coloring the spinodal unstable region in white, and marking the boundaries of the phase regions, where the transverse modes are activated and self-amplified, by the long and short dashed lines, respectively. Due to the very strong attractions, the spinodal unstable phase region is rather extensive. The phase region, where the transverse modes are activated, is limited at the central part of the spinodal unstable region, where the attractions among the quarks are more dominant. Under the phase region between the long and short dashed phase boundaries, both $\delta j_x$ and $\delta j_y$ oscillate with two real frequencies, and the dense clusters thus rotate spontaneously. The "tiny galaxies" thus form under such a phase region. The lower and higher real frequencies, as well as the growth rate, scaled by the wave number, are plotted in Fig. \ref{fig:Freqencies} as the functions of the temperature and the net charge density. As expected, both the lower frequencies ($\omega_1$) and the growth rates ($\Gamma$) are larger under the phase regions where the attractions are weaker, while the higher frequencies
($\omega_2$) are larger under the phase region where the attractions are stronger. 

Let us summarize the ingredients that are necessary for the presence of the spontaneous cluster rotation. First, we need a vector, or a tensor of higher order, mean-field potential for the appearance of the $\partial_t\mathbf j$ term, due to relativity, in the equations of motion. Second, we need attractions even stronger than those capable of rendering the spinodal instabilities, but not strong enough for rendering the self-amplification of the transverse modes. In particular, in the presence of a density dependent vector mean field potential, we require $U^\prime(\rho^\ast)$ varies in $(\chi^{\ast-1}_T,\chi^{-1}_{T\rm{min}})$. Under such a condition, the clusters should rotate with two angular frequencies, inversely proportional to the longitudinal size of the dense clusters in the classical limit, where the lower one increases, while the higher one decreases with $U^\prime(\rho^\ast)$. Quantum corrections on the dispersion relation, i.e., the wave number dependence of the frequencies, could be taken into account by using the linear response theory of quantum version~\cite{Li:2015pbv}. And a transport simulation, based on the equations of motion given by Eq. (\ref{eq:eom}), will be carried out in the near future for proving and visualizing the occurrence of the cluster rotation. Furthermore, the anomalous Lorentz transformation for the massive particles proposed in \cite{Li:2020vwh} will be employed as well, for investigating the spin in-homogeneity caused by the cluster rotation due to the spin vortical effect. Meanwhile, it needs to be mentioned that, the cluster rotation might not be the only source causing the local vortical flows, and hence the spin in-homogeneity, occurring in a spinodal phase separation. Small eddies might also emerge due to the friction between the dense and the dilute phase, which could be investigated in the transport simulations as well. All these simulations, en-lighted by this work, would build the connection between the spin in-homogeneity and the first order phase transition, and therefore open a new perspective for studying the phase transition from QGP to the hadronic matter.

\newpage
\bibliography{references}
\end{document}